\documentstyle[prb,aps,epsf]{revtex} \def\narrowtext{} \tighten \twocolumn
\input epsf.sty
\begin{document}
\draft

\title{Condensation Energy and Spectral Functions in High Temperature
Superconductors}
\author{
        M. R. Norman,$^1$
        M. Randeria,$^2$
        B. Jank\'o,$^1$
        and J. C. Campuzano$^{1,3}$
       }
\address{
         (1) Materials Sciences Division, Argonne National Laboratory,
             Argonne, IL 60439\\
         (2) Tata Institute of Fundamental Research, Mumbai 400005, 
             India\\
         (3) Department of Physics, University of Illinois at Chicago,
             Chicago, IL 60607\\
         }
\address{%
\begin{minipage}[t]{6.0in}
\begin{abstract}
If high temperature cuprate superconductivity is
due to electronic correlations, then the energy difference between
the normal and superconducting states can be expressed in terms of
the occupied part of the single particle spectral function. 
The latter can, in principle, be determined from 
angle resolved photoemission (ARPES) data.
As a consequence, the energy gain driving the development of the
superconducting state is intimately related to the dramatic changes in the
photoemission lineshape when going below $T_c$.
These points are illustrated in the context of the ``mode" model
used to fit ARPES data in the normal and superconducting states, where the
question of kinetic energy versus potential energy driven superconductivity is
explored in detail.
We use our findings to comment on the relation of ARPES data to the
condensation energy, and to various other experimental data.  In particular,
our results suggest that the nature of the superconducting transition is
strongly related to how
anomalous (non Fermi liquid like) the normal state spectral function is, and as
such, is dependent upon the doping level.
\typeout{polish abstract}
\end{abstract}
\end{minipage}}

\maketitle
\narrowtext

\section{Introduction}

The origin of high temperature superconductivity in the cuprates
is still a matter of great debate.  Recently, there have been several
different theoretical proposals for the mechanism of
high $T_c$ superconductivity, each of which leads to a 
characteristically different reason for the lowering of the free energy.
This has focused attention on
how various spectroscopic probes can yield information on the source 
of the condensation energy which drives the formation of the 
superconducting ground state.  

The first, and perhaps the most radical,
proposal is the interlayer tunneling theory of Anderson and co-workers,
where it is conjectured that the condensation energy is due
to a gain in the c-axis kinetic energy in the superconducting
state \cite{PWA}. Some measurements of the c-axis penetration
depth \cite{MOLER} are in conflict with the predictions of this theory.
Others, such as recent c-axis optical conductivity data \cite{BASOV} 
indicating a violation of the optical sum rule, are in support of this 
hypothesis, although alternative explanations have been proposed for these
observations \cite{IOFFE}.  An even more unusual suggestion has been recently
made by Hirsch and Marsiglio, \cite{HM} where they argue that the bulk of the
condensation energy comes from a gain in the in-plane kinetic energy.
A rather different approach proposes the lowering of the Coulomb
energy in the long wavelength, infrared region \cite{LEGGETT},
which has not been experimentally tested as yet.
A fourth approach advocates a lowering of the exchange energy in the 
superconducting state due to the formation of a resonant mode in the 
dynamic spin susceptibility \cite{SW,DZ} near ${\bf q} = (\pi,\pi,\pi)$, 
and has recently received 
experimental support from neutron scattering studies \cite{MOOK}.

We note that all of the above proposals focus on a part
of the Hamiltonian describing the system:
either a part of the kinetic energy, or a part of the interaction energy.
Correspondingly, the experiments to test these ideas focus on
two-particle correlation functions in 
a specific region of momentum and frequency space.

In this paper we propose to exploit a very general exact relation
between the one-particle Green's function of a system and 
its internal energy (see Eq.~1 below). 
This approach, in principle, allows us to 
determine the ``source'' of the condensation energy {\it without} making any
{\it a priori} assumptions about which piece of the Hamiltonian is
responsible for the gain in condensation energy.
The exact expression used involves moments of the occupied part of 
the one-electron spectral function, and since this quantity is directly 
related \cite{ARPES} to angle-resolved photoemission spectroscopy (ARPES) 
measurements, our approach also appears very promising from a 
practical point of view.

As a specific illustration of this general framework, we study
the condensation energy for a very simple
self-energy for the normal and superconducting
states which captures the essential features of the observed
ARPES lineshapes, the so-called mode model \cite{PRL97,ND,HUMP}. 
This analysis leads to several interesting
conclusions as discussed below, but most importantly,
it suggests an intimate connection between the nature of the normal
state spectral function (Fermi liquid or non Fermi liquid) and the
formation of sharply defined quasiparticle excitations below $T_c$,
and the gain in free energy in the superconducting state.

This paper is organized as follows.  In Section II, the formalism relating
the condensation energy to the spectral function is developed.  In Section III,
the mode model is introduced, and the nature of the resulting condensation
energy is discussed.
In Section IV, our observations concerning ARPES spectra are used to
comment on the results of previous spectroscopic studies, as well as the
origin of the condensation energy.  In Section V, we address the question
of the nature of the superconducting transition versus hole doping.
In Section VI, we offer some concluding remarks.  Finally, we include two
appendices.  Appendix A further explores questions raised in Section II in
regards to the full Hamiltonian and the virial theorem.
In Appendix B, we comment on the applicability of the
formalism of Section II to experimental data (ARPES and tunneling).

\section{Formalism}

We begin with the assumption that the condensation energy does not
have a component due to phonons, though as we mention below, this condition
can be relaxed.  We note that at optimal doping,
the isotope exponent $\alpha$ is essentially zero \cite{FRANCK}, 
and Chester \cite{CHESTER} proved that the change in ion kinetic energy 
between superconducting and normal states vanishes for $\alpha = 0$.  
To proceed, we assume an effective single-band Hamiltonian
which involves only two particle interactions.
Then, simply exploiting standard formulas \cite{FW,SS}
for the internal energy $U = \langle H - \mu N \rangle$ 
($\mu$ is the chemical potential, and $N$ the number of particles)
in terms of the one-particle Green's function, we obtain 
\begin{eqnarray}
\lefteqn{U_{N} - U_{S} =} \nonumber \\
& & \sum_{\bf k} \int_{-\infty}^{+\infty}d\omega 
(\omega + \epsilon_k) f(\omega)
\left[A_{N}({\bf k},\omega) - A_{S}({\bf k},\omega)\right]
\end{eqnarray}
Here and below the subscript $N$ stands for the normal state,
$S$ for the superconducting state. 
$A({\bf k},\omega)$ is the single-particle spectral function, 
$f(\omega)$ the Fermi function, and $\epsilon_k$
the bare energy dispersion which defines the kinetic energy part of 
the Hamiltonian.
Note that the $\mu N$ term has been absorbed into $\omega$
and $\epsilon_k$, that is, these quantities are defined relative to the
appropriate chemical potential, $\mu_N$ or $\mu_S$.  In general, $\mu_N$ and
$\mu_S$ will be different.  This difference has to be taken into
account, since the condensation energy is small.

The condensation energy is defined by the
zero temperature limit of $U_{N} - U_{S}$ in the above expression.
Note that this involves defining (or somehow extrapolating to) the
normal state spectral function at $T=0$. Such an
extrapolation, which we return
to below, is not specific to our approach, but required in all
estimates of the condensation energy.
We remark that Eq.~1 yields the correct condensation
energy, $N(0) \Delta^2/2$, for the BCS theory of 
superconductivity \cite{SCHRIEFFER}.

We also note that Eq.~1 can also be broken up into two pieces to
individually yield the thermal expectation value of the
kinetic energy (using $2\epsilon_k$ in the parentheses in front
of $f(\omega)$), and that of the potential energy
(using $\omega-\epsilon_k$ instead).
Further, this expression can also be generalized to the free energy
by including the entropy term as discussed by Wada \cite{WADA}.
Moreover, if the phonons can be treated in an harmonic approximation,
the terms missing in Eq.~1 (half the electron-phonon interaction, and all
other phonon terms) reduce to twice the phonon kinetic energy \cite{SS,WADA}.
The phonon kinetic energy can then be determined if the isotope
coefficient is known \cite{CHESTER}.  For $\alpha=1/2$, the missing terms
in this approximation reduce to twice the condensation energy, so that Eq.~1
is realized again, but with a {\it negative} sign.

The great advantage of Eq.~1 is that it involves just the occupied part of
the single particle spectral function, which is measured
by angle resolved photoemission spectroscopy (ARPES) \cite{ARPES}.
Therefore, in principle, one should be able to derive the condensation energy
from such data, if an appropriate extrapolation of the normal state spectral
function to T=0 can be made.
On the other hand, a disadvantage is that the bare energies, $\epsilon_k$, are
{\it a priori} unknown.  Note that these are not directly obtained from
the measured ARPES dispersion, which already includes many-body
renormalizations, nor are they simply determined by the
eigenvalues of a band calculation, as such calculations also 
include an effective potential term. 
Rather, they could be determined by projecting the kinetic
energy operator onto the single-band subspace.  Methodologies for doing this
when reducing to an effective single-band Hubbard model have been worked out
for the cuprates \cite{CDFT}, and could be exploited for this purpose.

Eq.~1 trivially reduces to the following:
\begin{eqnarray}
U_{N}-U_{S} = \sum_{\bf k} \epsilon_k
\left[n_{N}({\bf k})-n_{S}({\bf k})\right]
\nonumber \\
+ \int_{-\infty}^{+\infty} d \omega \omega f(\omega)
\left[N_{N}(\omega)-N_{S}(\omega)\right]
\end{eqnarray}
where $n({\bf k})$ is the momentum distribution function,
and $N(\omega)$ the single-particle density of states.
While ARPES has the advantage of giving information on both terms in this 
expression, other techniques could be exploited as well for the individual 
terms in Eq.~2.  For instance, $n({\bf k})$ in principle can be obtained
from positron annihilation or Compton scattering, while
$N(\omega)$ could be determined from tunneling data, although 
matrix elements could be a major complication for both tunneling and ARPES.

We conclude this Section with some remarks about a low-energy effective
single-band Hamiltonian used to derive Eq.~1 versus the 
{\it full} Hamiltonian
of the solid which includes quadratic dispersions for all (valence and core)
electrons, ionic kinetic energies, together with all Coulombic interactions
(see e.g., Ref.~\onlinecite{CHESTER}). 
As shown by Chester \cite{CHESTER} the full $H$ can be very useful
for studying the condensation energy. We discuss some points related to
such a description in the Appendix. 

Here we only wish to emphasize one important point which will
come up later in our analysis. In terms of the full Hamiltonian,
the transition to the superconducting state must be driven by
a gain in the potential energy (ignoring ion terms for this argument), 
as is intuitively obvious and also rigorously shown by Chester using
the virial theorem.  However, the kinetic energy terms in the
effective single-band Hamiltonian can (and in general do) incorporate
effects of the potential energy terms of the full Hamiltonian.
Further, there is no virial theorem restriction on the expectation
values of the kinetic and potential terms of the effective Hamiltonian
(since these do not, in general, obey the requisite homogeneity conditions).
As a consequence, there is nothing preventing the effective low-energy
Hamiltonian from having a superconducting transition driven by a
lowering of the (effective) kinetic energy.

\section{The Mode Model}

To illustrate the power of the formalism, as well as some of the
subtleties discussed above, we now analyze the condensation energy
arising from a spectral function described by a simple model self-energy
which captures some of the essential features of the ARPES data in the
important region of the Brillouin zone near $(\pi,0)$ in
the cuprates. These features are: 
(1) a broad normal state spectral function $A$ which seems 
$T$-independent in the normal state (except in the underdoped case, where
there is a pseudogap which fills in as $T$ increases), and thus can be 
used as the extrapolated ``normal'' state $A_N$ down to $T=0$ in Eq.~1. 
(2) A superconducting state spectral function $A_S$
which shows a gap, a sharp quasiparticle peak, and a dip-hump structure
at higher energies.
At a later stage, we will have to make some reasonable
assumptions about the ${\bf k}$-dependence of the spectral functions
to perform the zone sum in Eq.~1.

These nontrivial changes in the ARPES lineshape going from the
normal to the superconducting state have been attributed \cite{PRL97,ND} 
to the interaction of an electron with an electronic resonant mode 
below $T_c$, which itself arises self-consistently from the lineshape 
change. Strong arguments have been given which identify this resonant mode 
with one observed by magnetic neutron scattering \cite{PRL97,HUMP}. 
Thus our analysis below will also have bearing upon the
arguments mentioned in the Introduction which relate the resonant mode 
directly to changes in the exchange energy.

The simplest version of the resonant mode model 
is a self-energy of the form
\begin{equation}
\Sigma = \frac{\Gamma}{\pi}
{\rm ln}\left|\frac{\omega-\omega_0-\Delta}{\omega+\omega_0+\Delta}\right|
+i\Gamma\Theta(|\omega|-\omega_0-\Delta),
\end{equation}
where $\omega_0$ is the resonant mode energy, $\Delta$ the superconducting
energy gap, and $\Theta$ the step function. (A more complicated form 
has been presented in earlier work \cite{ND}.)
This self-energy is then used in the superconducting state spectral
function\cite{SCHRIEFFER}
\begin{equation}
A = \frac{1}{\pi}{\rm Im}
\frac{Z\omega+\epsilon}{Z^2(\omega^2-\Delta^2)-\epsilon^2}
\end{equation}
where $Z = 1-\Sigma/\omega$.  We note that for this form of $\Sigma$, the
spectral function $A_S$
will consist of two delta functions located at $\pm E$, where
$E$ satisfies two conditions: (1) it has a value less than $\omega_0+\Delta$
and (2) the denominator of Eq.~4 vanishes.  The weight of the delta
functions are then determined as \cite{MAHAN} $|dA^{-1}(\pm E)/d\omega|$.
In addition, there
are incoherent pieces for $|\omega|$ greater than $\omega_0+\Delta$.
We use the same self-energy for the (extrapolated) normal state 
with $\Delta=0$ and $\omega_0=0$, so that $A_N$ reduces to a Lorentzian
centered at $\epsilon$ with a full width half maximum of $2\Gamma$.

\begin{figure}
\epsfxsize=3.4in
\epsfbox{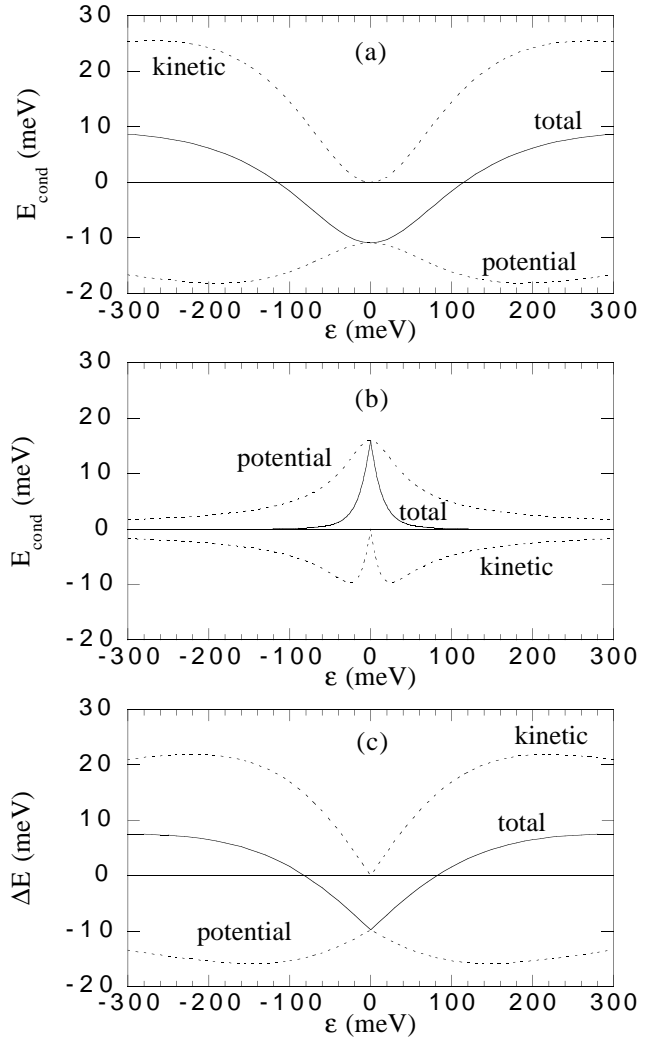}
\vspace{0.5cm}
\caption{(a) Condensation energy contribution, $E_{cond}$, versus single
particle
energy $\epsilon$ for the model self-energy of Eq.~3. As discussed in the
text, the quantity plotted is the result (as a function of $\epsilon$) after
the $\omega$-integration is done in Eq.~1.  
The parameters are $\Gamma$=230 meV,
$\Delta$=32 meV, and $\omega_0$=41.6 meV, which were obtained from fits
to ARPES spectra at $(\pi,0)$ \protect\cite{ND}.  The normal state is
obtained by setting $\omega_0$ and $\Delta$ to zero.  The dotted lines are
a decomposition of $E_{cond}$ into separate kinetic and potential energy pieces.
(b) Condensation energy contribution for the BCS theory using the same $\Delta$.
(c) A repeat of (a), {\it but} with the superconducting state replaced by the
normal state with $\omega_0$=41.6 meV
(and so labeled as $\Delta E$ instead of $E_{cond}$).}
\label{fig1}
\end{figure}

To begin with, for simplicity, we treat both 
$\omega_0$ and $\Delta$ as momentum independent.
It is straightforward to evaluate Eq.~1 with 
the sum over momentum reducing to an integral over $\epsilon$.  
In Fig.~1a, we plot the integrand of the $\epsilon$ integral 
(i.e., after the $\omega$ integral has been done).  The parameters used
are the same ones used earlier \cite{ND} to fit ARPES data near optimal
doping at the $(\pi,0)$ point.
The result is somewhat surprising.  
The integrand is negative for $\epsilon$ near zero (i.e.,
$k$ near $k_F$) and positive for $\epsilon$ far enough away.   This
should be contrasted with the BCS result \cite{TINKHAM}, shown in Fig.~1b,
where the
contribution at $k_F$ (which is $\Delta/2$) is maximal and positive.

To gain insight into this unusual result, we also show in Fig.~1a the
decomposition of this result into kinetic and potential energy pieces.
Unlike BCS theory (Fig.~1b), where the condensation is driven by the potential 
energy, in the mode model case, it is kinetic energy driven.  
To understand the unusual decrease in the kinetic energy as one goes
below $T_c$, we show in Fig.~2 the momentum distribution function 
$n({\bf k})$ plotted versus $\epsilon$.
Note that in contrast to BCS theory, $n({\bf k})$ is {\it sharper} in the
superconducting state than in the normal state.  The reason is very simple.
The (extrapolated) normal state is subject to a large broadening $\Gamma$ 
all the way down to $T=0$ which smears out $n({\bf k})$ on the
scale of $\Gamma$. At $T=0$ the result is simply:
$n_N({\bf k}) = 1/2 - \tan^{-1}\left(\epsilon/\Gamma\right)/\pi$.
In the superconducting state, although $\Delta$ broadens $n({\bf k})$
as in BCS theory, one now has quasiparticle peaks.  The effect of this
on sharpening $n({\bf k})$ is much larger than the broadening due to 
$\Delta$ (for $\Delta \ll \Gamma$), so the net effect is a significant 
sharpening. As a consequence, the kinetic
energy is lowered in the superconducting state.  

\begin{figure}
\epsfxsize=3.4in
\epsfbox{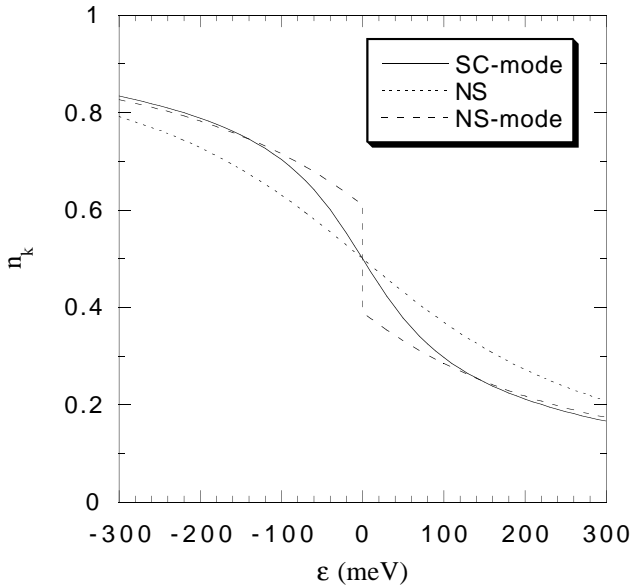}
\vspace{0.5cm}
\caption{Momentum distribution function versus $\epsilon$ in the 
superconducting
state (SC), normal state with $\omega_0$=0 (NS), and in the normal state 
with $\omega_0$=41.6 meV (NS-mode).  Same parameters as Fig.~1.}
\label{fig2}
\end{figure}

Note that these counterintuitive results would not have
been obtained had $\omega_0$ retained the same (non-zero) value in the normal
state.  In this case, sharp quasiparticles would
exist in the normal state, and all of our usual expectations
are fulfilled: $n_N({\bf k})$ 
would have had a step discontinuity (also illustrated in Fig.~2), 
and the normal state kinetic energy would have been considerably 
lower than the superconducting one.  In fact, for this situation, the model is
equivalent to that of Einstein phonons in an approximation where the
gap is treated as a (real) constant in frequency \cite{SCHRIEFFER}.
However, the normal state ARPES data near $(\pi,0)$ are clearly consistent with
$\omega_0 = 0$ and are $T$-independent with a $\Gamma \gg T$, which
suggests that the T=0 extrapolation used here is reasonable.

These points
are further illustrated in Fig.~1c, where we show the energy difference
between the normal state with $\omega_0$ non-zero, and the normal state
with $\omega_0$ zero.  Note the similarity to Fig.~1a, i.~e.~, the unusual
behavior in Fig.~1a is due to the formation of a gap in the incoherent part of
the spectral function, with the resulting appearance of quasiparticle states,
and thus not simply due to the presence of a superconducting energy
gap, $\Delta$.  Now, in the real system, it is the transition to a phase
coherent
superconducting state which leads to the appearance of the resonant mode
at non-zero energy, which causes the gap in $Im\Sigma$, which results in the
incoherent gap and quasiparticles, which in turn generates the mode.
Although this self-consistency loop clearly indicates the electron-electron
nature of the interaction (as opposed to an electron-phonon one),
the connection of these effects with the onset of phase
coherence (as opposed to the opening of a spectral gap, which is known to
occur at a higher temperature, $T^\ast$) is not understood at this time.
That is, the mode model is a crude simulation of the consequences of some
underlying microscopic theory which has yet to be developed.

As for the potential energy piece, we
note that the contribution to Eq.~1 at $k_F$ (where $\epsilon_k = 0$)
reduces to the first moment of the spectral function.  
In Fig.~3a, we plot the spectral function
at $k_F$ in both the normal and superconducting states. 
(For illustrative purposes, we have replaced the delta function peaks in 
the superconducting state by Lorentzians of half width half maximum 10 meV).
From this plot, we note that the quasiparticle peaks give a 
positive contribution to the condensation energy, but that at 
higher energies (large $|\omega|$), there is a negative contribution.  
This negative contribution is very important
because it is weighted by $\omega$ in the integrand of Eq.~1.  
To see this quantitatively, we plot in Fig.~3b the first moment 
difference at the Fermi surface ($\epsilon_k = 0$)
as a function of the lower cut-off on the $\omega$ integration (the
upper cutoff at $T=0$ is $\omega=0$).  
We clearly see the positive contribution due to
the quasiparticle peak and the (five times larger)
negative contribution due to the incoherent tail. 
This explains why the net contribution from the potential energy term 
is negative.  We can contrast this with BCS theory, where only the 
quasiparticle part exists, and so the net contribution is positive.

\begin{figure}
\epsfxsize=3.4in
\epsfbox{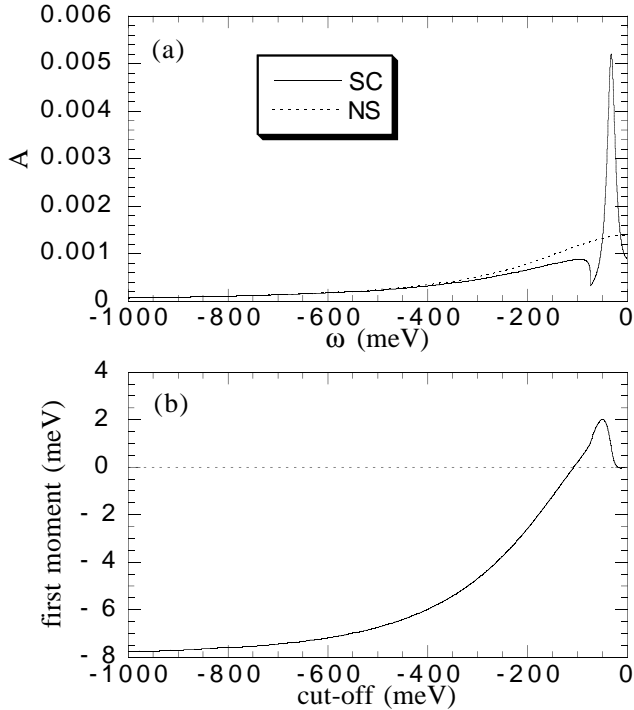}
\vspace{0.5cm}
\caption{(a) Spectral function at the Fermi surface 
($\epsilon = 0$) in the superconducting (SC) and normal states (NS).
(b) First moment contribution of (a) to the condensation energy
versus the lower cut-off in the $\omega$ integration in Eq.~1.  
Note positive contribution of the
quasiparticle peak, and large negative contribution from the high energy tail.
Same parameters as Fig.~1.}
\label{fig3}
\end{figure}

An interesting question concerns what happens in this model as the
broadening, $\Gamma$, is reduced.  In Fig.~4, we show results like for
Fig.~1a, but for various $\Gamma$ values.  As $\Gamma$ is reduced and
becomes comparable to $\Delta$, one crosses over from the unusual behavior
in Fig.~1a to a behavior very similar to that of BCS theory in Fig.~1b.
That is, the condensation energy crosses over from being kinetic energy
driven to being potential energy driven.  This is not a surprise, since
in the limit $\Gamma$ goes to zero, the model reduces to BCS theory.
The physics behind this, though, is quite interesting.  For large $\Gamma$,
the normal state is very non Fermi liquid like.  As $\Gamma$, is reduced,
though, the normal state becomes more Fermi liquid like \cite{GAMMA}.
As a consequence,
one crosses over from being kinetic energy driven to potential energy driven
(when $\Gamma \sim \Delta$).  The relation of kinetic energy driven behavior
with the presence of a non Fermi liquid normal state, and a Fermi liquid
superconducting state,
was realized early on by Anderson \cite{PWA3,PWA}, and will
be returned to again in Section IV of the paper.  Fig.~4 also draws
attention to the fact that being kinetic or potential energy driven is a
relative point.  Note in Fig.~4b that near $k_F$, the two contributions have
the {\it same} sign.  Individual terms, such as the potential energy in
Fig.~4b, and the kinetic energy in other cases we have explored,
can even change sign as a function of $\epsilon_k$.

\begin{figure}
\epsfxsize=3.4in
\epsfbox{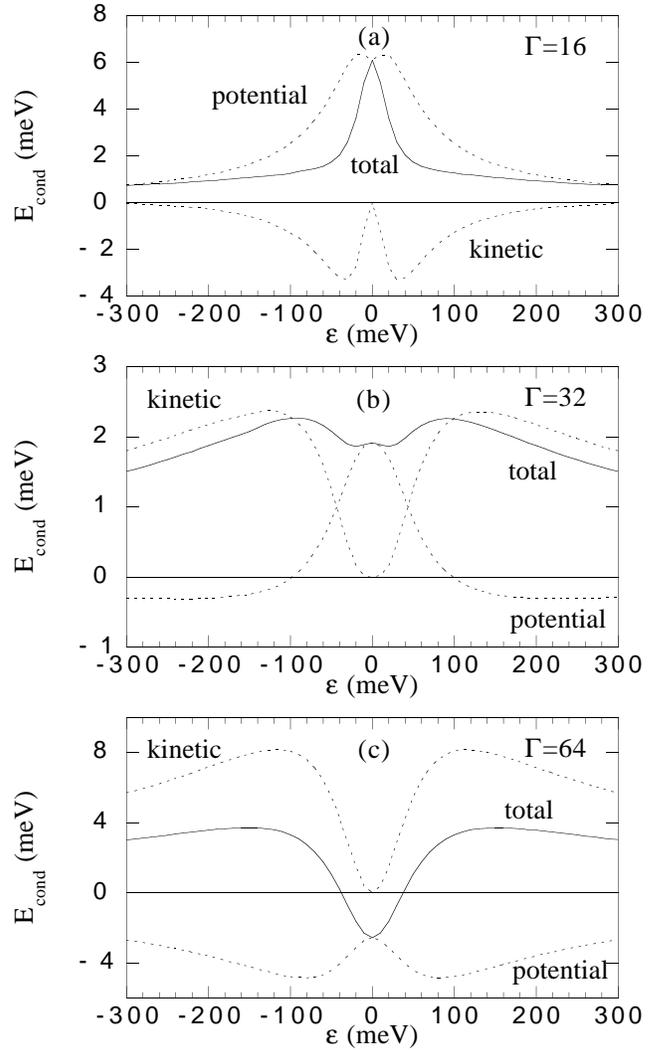}
\vspace{0.5cm}
\caption{Condensation energy contribution, as in Fig.~1a, for
(a) $\Gamma$=16 meV, (b) 32 meV, and (c) 64 meV.  $\Delta$ and $\omega_0$
are both 32 meV.  Note the crossover from kinetic energy driven behavior
to potential energy driven behavior as $\Gamma$ is reduced.}
\label{fig4}
\end{figure}

A potential worry which emerges from the above calculations is the large
contribution in Figs.~1 and 4 at large $|\epsilon|$.  In particular, in
most cases, the bulk of the contribution to the condensation energy comes
well away from the Fermi surface, in contrast to BCS theory.
In Fig.~1a, this is due to the large $\Gamma$,
which leads to a substantial rearrangement of the spectral function even for
large $|\epsilon|$, causing large contributions to both the potential and
kinetic
energy pieces.  Even in the case of Fig.~4a, where $\Gamma$ is quite small,
there is still a potential energy contribution at large $|\epsilon|$.  This can
be traced to the gap in the incoherent part of the spectral function, with the
resulting spectral weight being recovered around $\omega=\epsilon$, leading to
a potential energy shift.  Even in the BCS case, Fig.~1b, the individual
potential and kinetic energy pieces would not converge if integrated over an
infinite range in
$\epsilon$ (even though their difference would).  In BCS theory, this is
corrected by an ultraviolet cut-off (the Debye frequency).  We elect not to
include such a cut-off in the mode model, since it would be lead to another
adjustable parameter, and the $\epsilon$ integral is bound by the band edges,
and so is convergent.
In the real system, the ``mode" effects in the spectral function disappear
as one approaches the band edges, and as discussed in the following paragraph,
this effect can be crudely simulated by setting the mode energy proportional
to $\Delta_k$, the latter quantity in the d-wave case vanishing along the zone
diagonal where the band edges are located.

Although we plot only the differences in Figs.~1, ~3b, and 4,
the individual normal and
superconducting state terms are quite large.  This raises the question of
what the value of Eq.~1 would actually be if summed over the zone.
To do this, we must make some assumptions about what the momentum dependence
of various quantities are.  For simplicity sake, we will
treat $\Gamma$ as $k$ independent, though we note that available ARPES data 
are consistent with this quantity being reduced in size as one moves from 
$(\pi,0)$ towards the Fermi crossing along the $(\pi,\pi)$ direction.  
In the first sum, denoted by case (a), we
treat $\Delta$ and $\omega_0$ as $k$ independent.  
In the second sum, denoted by case (b), we replace
$\Delta$ by $\Delta_k=\Delta_0(\cos(k_xa) -\cos(k_ya))/2$, where  $\Delta_k$
is the standard d-wave gap function, but still retain
a $k$-independent $\omega_0$.  In the third sum (c), in addition
to the d-wave $\Delta_k$ we also take 
$\omega_0=c|\Delta_k|$, with the $k$
dependence of $\omega_0$ crudely simulating the fact that the mode effects
in the spectral function are reduced as one moves away from the $(\pi,0)$
points of the zone \cite{ND}.  The values of these parameters are the same
as used in Fig.~1a, and are
consistent with ARPES and neutron data for Bi2212 ($\Gamma$=230 meV,
$\Delta_0$=32 meV, $c$=1.3).  To perform the zone sum, we have to make
some assumptions on what the $\epsilon_k$ are.  As the mode model is designed
to account for the difference between the normal state and superconducting
state, we elect to use normal state ARPES dispersions for
$\epsilon_k$ \cite{NORM95}, though we caution that this represents a
different choice for the ``kinetic" energy part of the effective single-band
Hamiltonian than is typically used \cite{CHOICE}.  Because this dispersion has
particle-hole
asymmetry, the chemical potential will not be the same in the superconducting
state as in the normal state.  The chemical potential is thus tuned to
achieve the same density (a hole doping x=0.16) as the normal state.  Note
that the normal state density itself is a function of $\Gamma$ (we assume
$\omega_0$=0 for the normal state).

Performing the zone sum, we find condensation energies of 
+3.6, +3.3, and +1.1 meV, per CuO plane, for case (a), (b) and (c)
respectively. We note that the last result is the more
physically appropriate, and though small, is somewhat larger than
the condensation energy of 1/4 meV per plane estimated by Loram {\it et al}
from specific heat data for optimal doped YBCO \cite{LORAM}.
The above values will be reduced if a more realistic $k,\omega$
dependence is used for $\Gamma$, since, as we noted above, $\Gamma$
decreases as one moves away from $(\pi,0)$.
As consistent with Fig.~1a, the contribution to the condensation energy 
is negative for an anisotropic shell around the Fermi surface 
(due to the anisotropy of $\Delta_k$ and $\epsilon_k$),
and positive outside of this shell.  Again, this will be sensitive to the
$k$ dependence of $\Gamma$, as can be seen from Fig.~4.
We also remark that there are chemical potential shifts of +2.6, +2.1, and
+1.4 meV, respectively, for case (a), (b) and (c).  Again, the last value is the
more physically appropriate.  It is very interesting to note that somewhat
smaller positive shifts (around +0.6 meV) have been seen experimentally in
YBCO \cite{VDM}.  These shifts are a consequence of particle-hole
asymmetry and the change in $n_k$ when going into the superconducting state.

\section{Connections With Previous Work}

While a quantitative evaluation of Eq.~1 using experimental
data as input on the right hand side must await further progress
as discussed in Appendix B,
several qualitative points can be made even at this stage. From 
Eq.~1, there is a one to one correspondence between the changes in
the spectral function and the condensation energy.  That is, the condensation
energy is due to the profound change in lineshape seen in photoemission
data when going below $T_c$.  When summed over the zone, this in turn
leads to changes in the tunneling density of states (second part of Eq.~2). 
These spectral function changes cause, and are themselves caused by,
changes of various two particle correlation functions, such as
the optical conductivity and the dynamic spin susceptibility, which
have previously been used by others to comment about the nature of the
condensation energy.

In this context, we now discuss the earlier work concerning the c-axis
conductivity.  The most dramatic changes in the ARPES lineshape when going below
$T_c$ occur near the $(\pi,0)$ points of the zone.  It is exactly these points
of the zone which appear to have the largest c-axis tunneling matrix elements
associated with them \cite{OKA}. 
Previous work has found a strong correlation between
the c-axis conductivity and ARPES spectra near the $(\pi,0)$ points of the
zone \cite{CARDONA,IOFFE}.  Therefore, it is rather straightforward to
speculate that it is the formation of strong quasiparticle peaks in these
regions of the zone, and the resulting changes in the spectral function at
higher binding energy, which is responsible for the lowering of the c-axis
kinetic energy.  We note that earlier, Anderson \cite{PWA3,PWA} had remarked
that if the quasiparticle weight is coming from high binding energy, then one
would expect a lowering of the kinetic energy.  This in fact is what is
occuring in the mode model calculations, though we note from our work that
the true quantity which determines the sign of the kinetic energy change in
the vicinity of $k_F$ is the gradient of the momentum distribution function
at $k_F$.

We also remark that the change in c-axis kinetic energy has been recently
addressed by Ioffe and Millis in
the context of the same mode model used in the current paper \cite{IOFFE}.
These effects would enter directly in Eq.~1 by including a c-axis tunneling
contribution to $\epsilon_k$ \cite{IOFFE}.  As for the in-plane kinetic energy,
it is so large that it is difficult to determine its contribution to Eq.~1
from optical conductivity data because of some of the same normalization
concerns mentioned in Appendix B in regards
to ARPES and tunneling data.  Still, if the mode model calculation is a
reflection of reality, we can speculate that $n({\bf k})$ will
probably sharpen in the superconducting state, leading to a lowering of
the in-plane kinetic energy.  How large the effect will be is somewhat
difficult to determine, in that the same regions of the zone where large
changes are seen in the ARPES lineshape are also characterized by small
Fermi velocities (the optical conductivity involves a zone sum weighted by
$v_F^2$).  Along the $(\pi,\pi)$ direction, for instance, there is still
some controversy concerning how dramatic the lineshape change is below
$T_c$ \cite{KAMINSKI,VALLA}.
Also, as can be seen from Fig.~4, this question is very dependent
on the variation of the normal state lineshape in the zone.  Although
the lineshape near $(\pi,0)$ is highly non Fermi liquid like, the behavior
along the $(\pi,\pi)$ direction appears to be marginal Fermi liquid
like \cite{KAMINSKI,VALLA}.  As remarked in Section III, the more Fermi
liquid like the normal state lineshape is, the greater the tendency is to
switch over to potential energy driven behavior instead.
Improved experimentation should again lead to a resolution of these issues.

This brings us to the question concerning the relation of the magnetic resonant
mode observed by neutron scattering to the condensation energy.
All calculations of the resonant mode assume the
existence of quasiparticle peaks.  In the absence of such quasiparticle
peaks, a sharp resonance is not expected.  That is, the sharp resonance
observed by neutron scattering, and the resulting lowering in the exchange
energy part of the t-J Hamiltonian, is again a consequence of the formation of
quasiparticle states.  In this context, it is important to note that the
d-wave coherence factors associated with quasiparticle states are important
for the formation of the resonance, whether in the context of
calculations in the particle-hole channel \cite{FONG}, or in the 
particle-particle scenario proposed by Demler and Zhang \cite{PRL95}.
In any case, this again supports our statement, motivated by Eq.~1, 
that it is the dramatic change in the ARPES spectra below $T_c$ which is
the source of the condensation energy.

In this regard,
we note a puzzling feature in connection with the mode model.  Although it
was designed to take into account the effect of the magnetic resonance mode
on the spectral function, the condensation in the mode model is 
kinetic energy driven.  This is in contrast to the potential energy driven 
nature of the condensation with the resonant mode discussed in the 
context of the t-J model \cite{SW,DZ,MOOK}, despite the same underlying
physics.  There are two possibilities for this apparent discrepancy.
First, the break-up of the Hamiltonian into
potential and kinetic energy pieces depends on the particular single-band
reduction which is done.  The superexchange energy, which is a
kinetic energy effect at the level of the Hubbard model \cite{PWA2},
appears as a potential energy term when reduced to the t-J Hamiltonian.  
In the mode model, the kinetic energy is equated to $\epsilon_k$ based on 
normal state ARPES dispersions \cite{CHOICE}, while the potential energy term
leads to effects described by the $\Sigma$ of Eq.~3.

The second possibility is
that the argument of Ref.~\onlinecite{DZ} is confined to low energies of
order $\Delta$.  
As demonstrated in Fig.~3b, if the mode model is confined
to such energy scales, the first moment (i.e., the potential) term would
reverse sign, since the
quasiparticle peak always gives a positive contribution to the first moment.
That is, one would expect the resonance to lower the
exchange energy since it is a consequence of the quasiparticle states, which
lower the potential energy in Eq.~1.  It is the difference in the high energy
incoherent tails (Fig.~3), though, which is ultimately responsible for the
increase of the net potential energy in Fig.~1a.  This would imply
that the neutron scattering results \cite{MOOK} may change if more complete
data at
higher energies and other $q$ values are obtained.  That is, the true
answer will depend on where the weight for the neutron resonance is coming
from, in complete analogy to the earlier mentioned argument of
Anderson \cite{PWA3} in regards to where the quasiparticle weight is coming
from.

This discussion again emphasizes that the current debate concerning
kinetic energy driven superconductivity versus potential energy driven
superconductivity must be kept in proper context, as the very definition of 
the kinetic and potential pieces is dependent upon what effective low
energy Hamiltonian one employs, and what energy range one considers.

\section{Doping Dependence}

The condensation energy as
estimated from specific heat is known to decrease strongly as the doping is
reduced \cite{LORAM}.  This is despite the increase of the spectral
gap \cite{HARRIS,JOHNZ,HUMP}.  There are two reasons for this suggested by the
above line of reasoning.  First, the normal state itself at $T_c$ already
exhibits a large spectral gap, the so-called pseudogap, which acts to reduce 
the difference in Eq.~1.  
Second, the weight of the quasiparticle peak strongly
decreases as the doping is reduced \cite{HUMP}.  
This reduces the quasiparticle
contribution to both the first moment and to $n({\bf k})$.
We caution that the normal state extrapolation down 
to T=0 will be more difficult to estimate for underdoped experimental 
data because of the influence of the pseudogap, 
which is known to fill in as a function of temperature \cite{NATURE2}.
Still, the available underdoped ARPES and tunneling data are certainly in
support of a smaller condensation energy than overdoped data due to the 
pseudogap, which is in agreement with conclusions based on specific heat 
data \cite{LORAM2}.
The new contribution to these arguments is the strong reduction of the
weight of the quasiparticle peak in the underdoped case which makes the
condensation energy smaller still.  In fact, based on our arguments, the
strong reduction of the superfluid density upon underdoping is almost
certainly connected with the strong reduction in the quasiparticle weight.

Finally, Anderson \cite{PWA4} has
speculated that the superconducting transition temperature is potential
energy driven on the overdoped side, kinetic energy driven on the underdoped
side.  This is a distinct possibility, since
$\Gamma$ is known from ARPES data \cite{KAMINSKI} to be strongly reduced as
the doping increases on the overdoped side, and as Fig.~4 demonstrates, one
might expect (if the mode model is a reflection of reality) a crossover from
kinetic energy driven behavior to potential energy
driven behavior as $\Gamma$ is reduced.  In this context, we note the Basov
{\it et al}
result \cite{BASOV} that the lowering of the c-axis kinetic energy appears
to be confined to the underdoped side of the phase diagram.
Moreover, if one attributes $T^\ast$ on the underdoped
side to the onset of pairing correlations \cite{VARENNA}, 
then one anticipates a potential energy gain 
due to pairing to occur at this finite temperature crossover.
At $T_c$, phase coherence in the pair field is established, and the resulting
quasiparticle formation \cite{PHENOM} and related spectral changes could
lead to a kinetic energy driven transition of the sort discussed above.
We emphasize ``could", since in the context of Eq.~1, there is no unambiguous
evidence yet from real ARPES data that such is the case.

\section{Concluding Remarks}

We conclude this paper by noting that the above arguments based on
condensation energy considerations highlights one of the key question
of the high $T_c$ problem: why do quasiparticle
peaks only appear below $T_c$?  This is especially relevant in the underdoped
case, since the spectral gap turns on at a considerably higher temperature
than $T_c$, but the quasiparticle peaks again form only at $T_c$ \cite{PHENOM}.
This implies that there is a deep connection between the onset of
phase coherence in the pair field and the onset of coherence in the single
electron degrees of freedom \cite{PWA}.
We suggest that the understanding of
this connection will be central to solving the high $T_c$ problem.
The result of the current paper is that Eq.~1 
brings this issue into much shaper focus.
In particular, as a cautionary note, the incoherent part of the spectral
function is likely to be as important as the
quasiparticle component in determining the condensation energy (Fig.~3).
That is, it is the overall shape of the spectral function (the peak-dip-hump
behavior of Fig.~3a),
rather than just the quasiparticle part, which is ultimately responsible
for the total condensation energy.
We believe that experimental data analyzed in the context of Eq.~1
will play an important role in providing a solution to the high $T_c$ problem.

\acknowledgments

We thank Hong Ding, Helen Fretwell, Adam Kaminski, and Joel Mesot for
discussions concerning the ARPES data, and Laura Greene, Christophe Renner, and
John Zasadzinski for providing their tunneling data.
This work was supported by the the U. S. Dept. of Energy,
Basic Energy Sciences, under contract W-31-109-ENG-38, the National
Science Foundation DMR 9624048, and DMR 91-20000 through the Science and
Technology Center for Superconductivity. MR is supported in part by the
Indian DST through the Swarnajayanti scheme.

\appendix
\section{The Full Hamiltonian and the Virial Theorem}

In this Appendix, we make further comments on some issues
which were briefly discussed at the end of Section II,
relating to the use of the full Hamiltonian versus an effective single-band
Hamiltonian.

We note that as written, Eq.~1 does not apply to the {\it full} Hamiltonian
of the solid which includes all the electronic and ionic degrees of freedom
together with their Coulombic interactions as discussed in
Ref.~\onlinecite{CHESTER}.
In principle an expression similar to Eq.~1 could be written
if the quantities in Eq.~1 were replaced by matrices in reciprocal lattice
space \cite{HEDIN}.  For our purposes, where
an energy difference is being looked at, a unitary transformation to band
index space would be desirable. The resulting off-diagonal terms would then
represent interband transitions. These could be of potential importance,
even for the energy difference. For example, the violation of the
c-axis optical conductivity sum rule \cite{BASOV} implies
a change in interband terms so that the total optical sum rule is satisfied.

The usefulness of the full Hamiltonian is that
one can use the virial theorem \cite{CHESTER,LP}
$2K-nV-3P\Omega=0$, exploiting the fact that the
kinetic energy $K$ is a homogeneous function of order
2 in momentum, and the potential energy $V$ is a homogeneous function
of order n in position. Here $P$ is the pressure and $\Omega$
denotes the volume. For Coulomb forces $n=-1$, and ignoring the pressure terms
(which are negligible at ambient pressure), this reduces to $2K+V=0$.

If we assume that the form of Eq.~1
applies to the full Hamiltonian (which could be possible
if all interband terms dropped out of
the energy difference, as well as all electron-ion and ion-ion terms)
then by using the virial theorem, the right hand side of Eq.~2
can be shown to reduce to 2/3 the first
moment of the density of states at $T=0$.
In addition, the change in the kinetic energy would be
the negative of the condensation energy, with the potential energy twice the
condensation energy.

This reduced form of Eq.~2, though, must be treated with 
extreme caution, and is likely not useful to the problem at hand.
The reason is that the kinetic energy
and potential energy terms of the full Hamiltonian are not the same as the
kinetic and potential energy terms of the effective single-band Hamiltonian.  
It is only for the former that the virial theorem manipulations would be 
allowed.  As an example, BCS theory obeys Eq.~2, but not the reduced form.

\section{Comments on ARPES and Tunneling}

The purpose of Section III was to demonstrate how Eq.~1 works out in
practice for a model where exact calculations could be done.  This is important
when considering real experimental data.  We have
spent considerable effort analyzing Eqs.~1 and 2
using experimental data from ARPES and tunneling
as input, and plan to report on these endeavors in a future
publication.  
But given what we have learned from the mode model, some of the problems
associated with an analysis based on experimental data
can be appreciated.  First, the condensation energy is obtained by 
subtracting two large numbers.  Therefore, normalization
of the data becomes a central concern.  Problems in this regard when 
considering $n({\bf k})$, which is the zeroth moment of the ARPES data,
were discussed in a previous experimental paper \cite{SHIFT}.
For the first moment, these problems are further amplified
due to the $\omega$ weighting in the integrand.  This can be appreciated from
Fig.~3, where the bulk of the contribution in the mode model comes from the
mismatch in the high energy tails of the normal state and superconducting 
state spectral functions.  When analyzing
real data, we have found that the tail contribution, either from ARPES or from
tunneling, is very sensitive to how the data are normalized.  Different choices
of normalization can even lead to changes in sign of the first moment.

Another concern concerns the ${\bf k}$ sum in Eq.~1. 
Both ARPES and tunneling have (their own distinct) ${\bf k}$-dependent 
matrix elements, which lead to weighting
factors not present in Eq.~1.  For ARPES, these effects can in principle be
factored out by either theoretical estimates of the matrix 
elements \cite{BANSIL},
or by comparing data at different photon energies to obtain information on
them \cite{JOEL}.  
For tunneling, information on matrix elements can be obtained by
comparing different types of tunneling (STM, tunnel junction, point contact),
or by employing directional tunneling methods.

Another issue in connection with experimental data is an appropriate
extrapolation of the normal state to zero temperature.  Information on this
can be obtained by analyzing the temperature dependence of the normal state
data, remembering that the Fermi function will cause a temperature dependence
of the data which should be factored out before attempting the $T=0$
extrapolation.  We finally note that the temperature dependence issue is
strongly coupled to the normalization problem mentioned above.  In ARPES, the
absolute intensity can change due to temperature dependent changes in absorbed
gasses, surface doping level, and sample location \cite{SHIFT}.  In tunneling,
the absolute
conductance can change due to temperature dependent changes in junction
characteristics.
In both cases, changes of background emission with temperature is
another potential problem.

Despite these concerns,
we believe that with careful experimentation, many of these difficulties
can be overcome, and even if an exact determination of Eq.~1 is not possible,
insights into the origin of the condensation energy will certainly be
forthcoming from the data.  This is particularly true for ARPES, which has
the advantage of being ${\bf k}$ resolved and thus giving one information on the
relative contribution of different ${\bf k}$ vectors to the condensation energy.


\begin{references}
\bibitem{GAMMA}
For the purposes of this paper, by a Fermi liquid, we mean a quasiparticle
peak which is sharp compared to its energy position.  We note that the
normal state $Im\Sigma$ in the mode model is a constant, and thus
does not vanish as $\omega^2$.
\bibitem{CHOICE}
In the mode model, the normal state self-energy is an imaginary constant,
so the normal state dispersion reduces to that from the effective kinetic
energy term.  This would not be true in a more sophisticated model which had
a frequency dependent self-energy.
\end{references}
\end{document}